\documentclass{ptapap}

\usepackage{graphicx}
\usepackage{caption}
\usepackage{subcaption}
\usepackage{rotating}

\author{Jakub Klencki}[OA]
\author{{\L}ukasz Wyrzykowski}[OA]
\affil[OA]{Warsaw University Astronomical Observatory\\
  Al. Ujazdowskie 4, 00--478 Warsaw, Poland}

\headtitle{Real-time detection of transients in OGLE-IV\dots}
\title{Real-time detection of transients in OGLE-IV with application of machine learning}

\begin{document}

\maketitle

\begin{abstract}

The current bottleneck of transient detection in most surveys is the problem of rejecting numerous artifacts from detected candidates. 
We present a triple-stage hierarchical machine learning system for automated artifact filtering in difference imaging, based on self-organizing maps. The classifier, when tested 
on the OGLE-IV Transient Detection System, accepts $~97\%$ of real transients while removing up to $~97.5\%$ of artifacts.

\end{abstract}

\section{Introduction}

The task of analyzing the observational data is especially challenging for time-domain surveys searching for transients (such as OGLE \citep{Udalski2015}), 
as they usually attempt to perform data analysis as close to real-time as possible.
Detecting an ongoing transient as early as possible not only allows for triggering extra follow-up observations, leading to far better light curve coverage, 
but also gives a chance to observe the earliest stages of the event - possibly essential for understanding the physics of the phenomena.
Machine learning techniques are a promising approach to help solve this problem by performing data classification automatically \citep{B&R2012}
In our work we aim to further develop the OGLE-IV Transient Detection System \citep{Wyrzykowski2014} by automatically rejecting numerous bogus detections -
the most time consuming stage of transient detection. \newline

\begin{figure}[htb]
        \centering
        \begin{subfigure}[b]{0.23\textwidth}
                \includegraphics[width=\textwidth,natwidth=832,natheight=832]{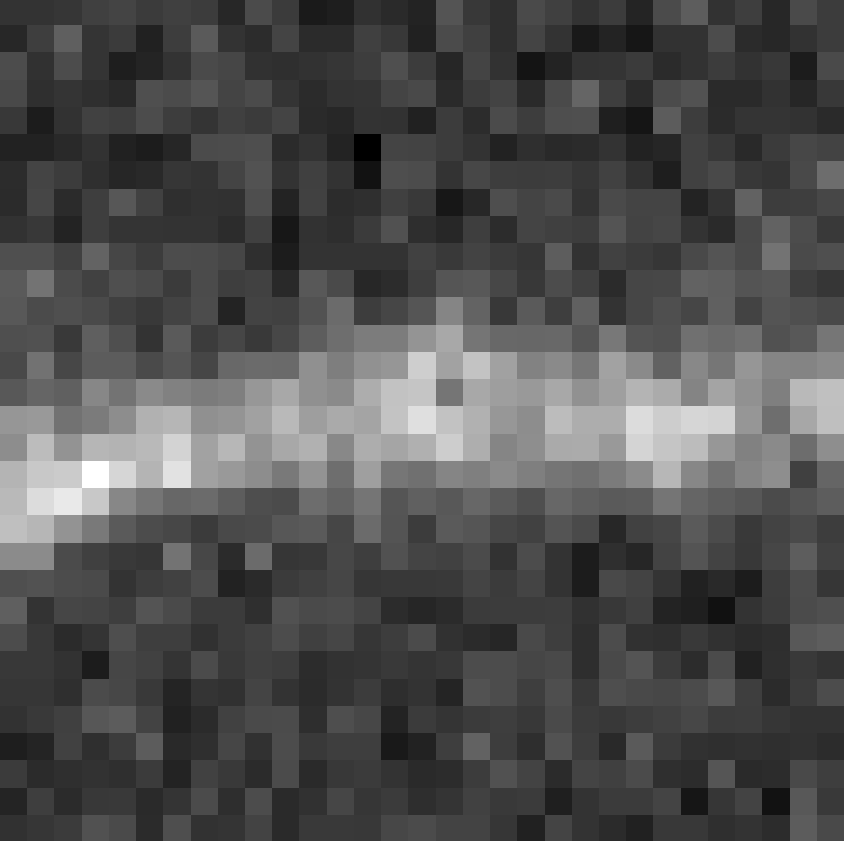}
                \caption{satellite's crossing}
        \end{subfigure}%
        ~ %add desired spacing between images, e. g. ~, \quad, \qquad, \hfill etc.
        \begin{subfigure}[b]{0.23\textwidth}
                \includegraphics[width=\textwidth,natwidth=832,natheight=832]{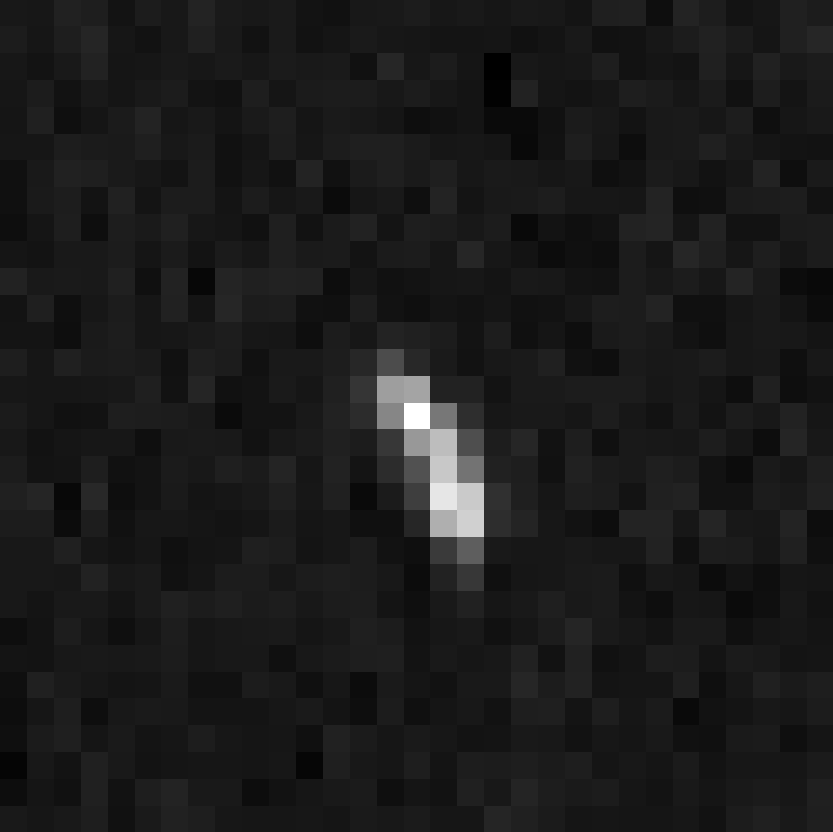}
                \caption{cosmic ray \newline}
        \end{subfigure}
        ~ %add desired spacing between images, e. g. ~, \quad, \qquad, \hfill etc.
        \begin{subfigure}[b]{0.23\textwidth}
                \includegraphics[width=\textwidth,natwidth=832,natheight=832]{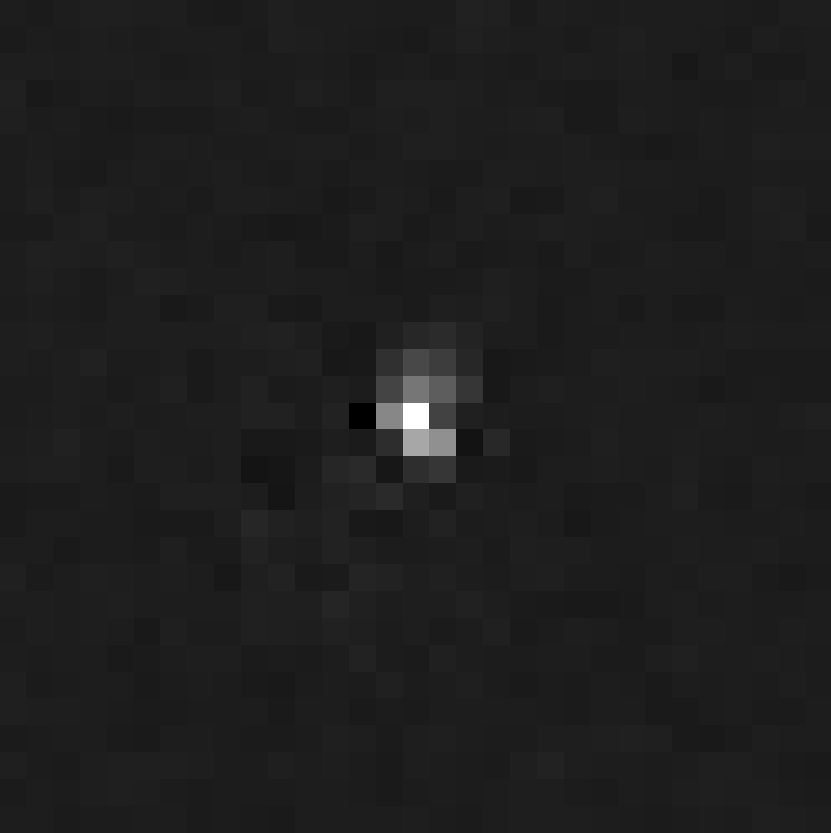}
                \caption{subtraction failure, too faint}
        \end{subfigure}
        ~ %add desired spacing between images, e. g. ~, \quad, \qquad, \hfill etc.
        \begin{subfigure}[b]{0.23\textwidth}
                \includegraphics[width=\textwidth,natwidth=832,natheight=832]{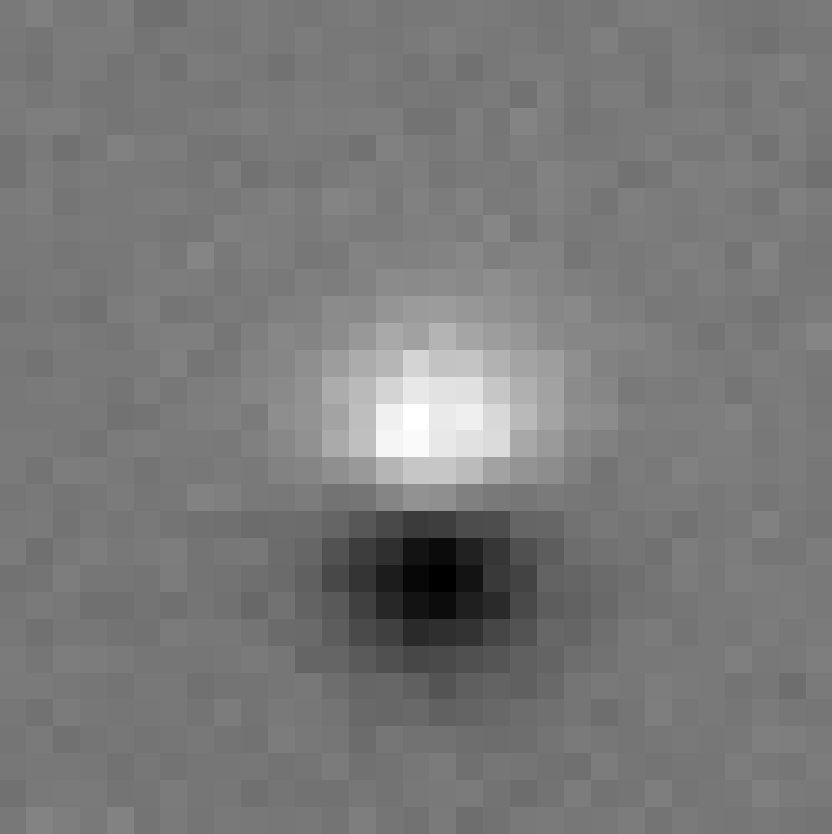}
                \caption{high proper motion stellar object}
        \end{subfigure}
        \caption{Most common types of artifacts from the OGLE transient pipeline. Presented as 31x31 pixel cutouts from image differences. }
\end{figure}

The OGLE pipeline identifies candidates for transients using Difference Imaging Analysis and cuts the subtraction images into
thumbnails centered around positive subtraction residuals. The pipeline
produces thousands of candidates each night, the vast majority of which being unwanted artifacts and false
subtractions – we call them bogus detections (see Fig.1).

\section{Method: Self-Organizing Maps}
Self-Organizing Maps (SOM) \citep{Kohonen1982} is an artificial neural network, usually in a form of 2D
array of nodes with weight vectors associated to them. Each analyzed datum is
represented in a form of a numeric vector and mapped onto node with weight vector most similar to it – the winner node. \\

The SOM organizes itself during an unsupervised learning process. For each training vector \textbf{v} the winner
weight vector \textbf{w} is found. Then the winner and all his neighbouring nodes are adjusted with a learning rate
$\alpha$, which is decreasing from 1 to 0 during the training process:

\begin{equation}
 w_{new} = w_{old} + \alpha (v - w_{old})
\end{equation}

\begin{figure}[htb]
        \centering
        \begin{subfigure}[b]{0.27\textwidth}
                \includegraphics[width=\textwidth,natwidth=832,natheight=832]{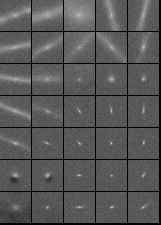}
                \caption{first stage: SOM trained on dataset
containing mostly cosmic ray and
satellite crossing types of bogus
detection.}
        \end{subfigure}%
        ~ %add desired spacing between images, e. g. ~, \quad, \qquad, \hfill etc.
        \begin{subfigure}[b]{0.27\textwidth}
                \includegraphics[width=\textwidth,natwidth=832,natheight=832]{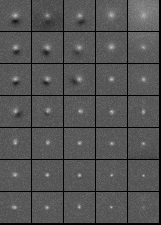}
                \caption{second stage : SOM trained on dataset of
candidates appearing on at least 2
independent observations }
        \end{subfigure}
        ~ %add desired spacing between images, e. g. ~, \quad, \qquad, \hfill etc.
        \begin{subfigure}[b]{0.27\textwidth}
                \includegraphics[width=\textwidth,natwidth=832,natheight=832]{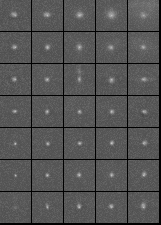}
                \caption{third stage: SOM trained on dataset
containing candidates which had
been accepted by second-stage
SOM (b).}
        \end{subfigure}
        \caption{Visualizations of trained SOMs}
\end{figure}
\section{Results}
Using representative training datasets from OGLE-IV pipeline we trained three different SOMs to be used at
subsequent stages of candidates verification, resulting altogether in a triple-stage verifier (Fig.2).
Thumbnail images of candidates are being represented as vectors of pixel and
their histogram values. In order to use SOM as a verification tool for transient candidates one has to choose which nodes form the
acceptance region (i.e. images classified into them are verified as positives and accepted). The best to
choose are nodes with highest percentage of true positives among all classified objects – we call it node's
purity, similary to \citep{Bloom2012}.

\begin{figure}[htb]
        \centering
        \begin{subfigure}[b]{0.25\textwidth}
                \includegraphics[width=\textwidth,natwidth=832,natheight=832]{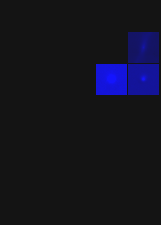}
                \caption{first stage SOM}
        \end{subfigure}%
        ~ %add desired spacing between images, e. g. ~, \quad, \qquad, \hfill etc.
        \begin{subfigure}[b]{0.25\textwidth}
                \includegraphics[width=\textwidth,natwidth=832,natheight=832]{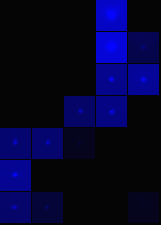}
                \caption{second stage SOM}
        \end{subfigure}
        ~ %add desired spacing between images, e. g. ~, \quad, \qquad, \hfill etc.
        \begin{subfigure}[b]{0.25\textwidth}
                \includegraphics[width=\textwidth,natwidth=832,natheight=832]{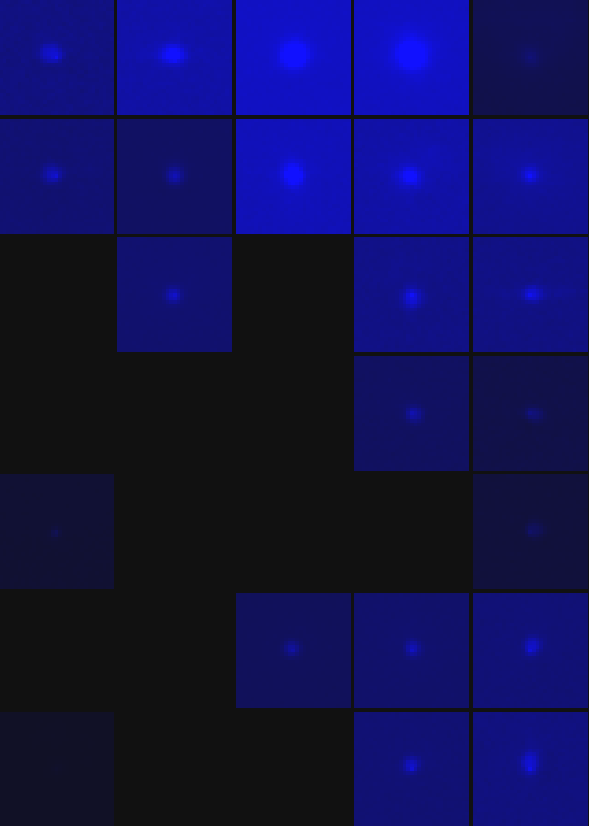}
                \caption{third stage SOM}
        \end{subfigure}
        \begin{subfigure}[b]{0.075\textwidth}
                \includegraphics[width=\textwidth,natwidth=832,natheight=832]{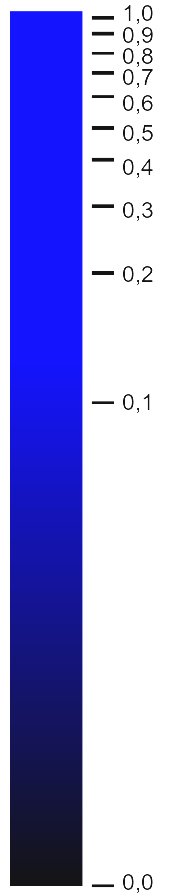}
        \end{subfigure}
        \caption{Relative purity of nodes for each SOM, estimated based on test verifications of candidates of known types
        (real or bogus).}
\end{figure}

\begin{figure}[htb]
\centering
\includegraphics[width=0.7\textwidth,natwidth=832,natheight=832]{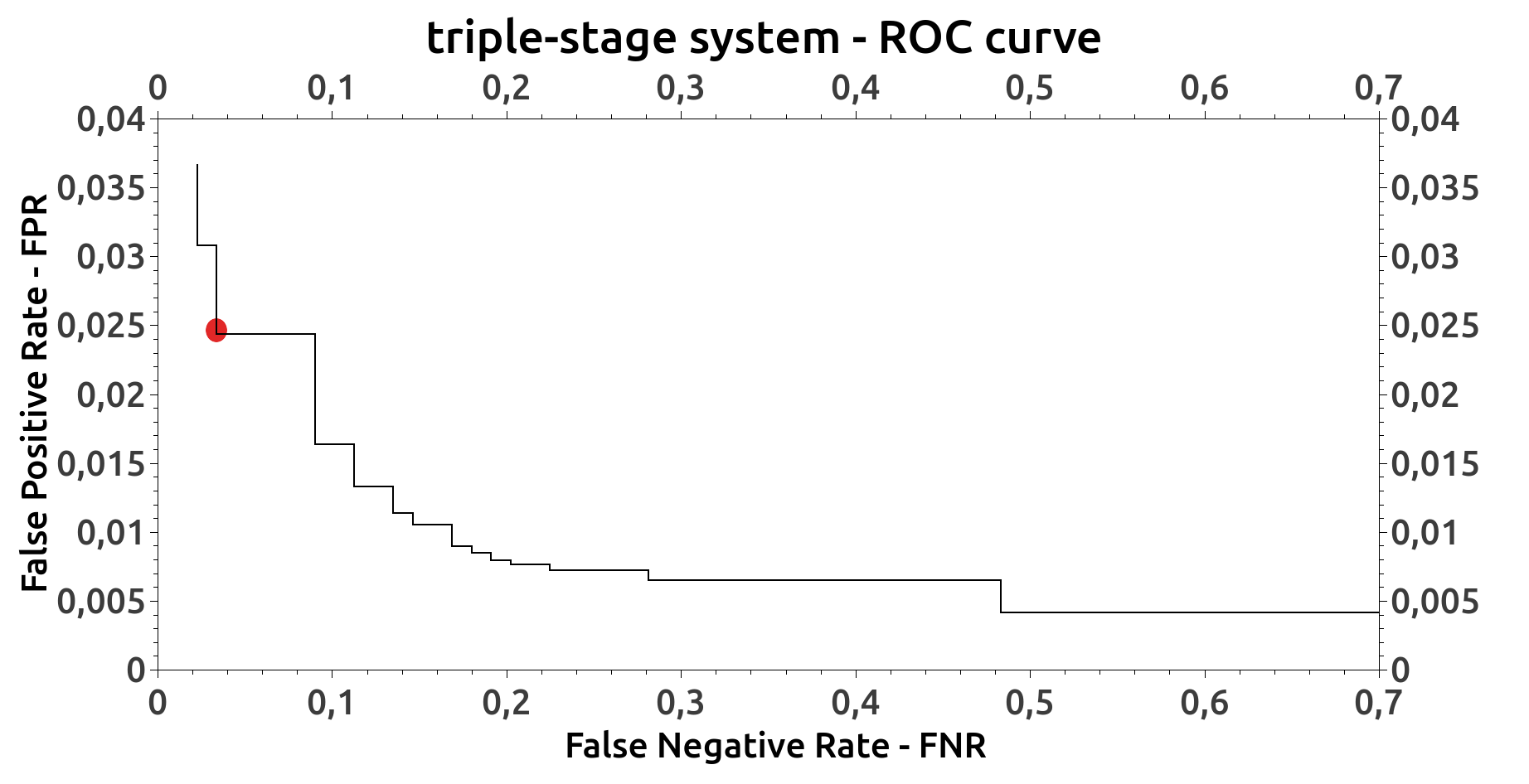}
\caption{Receiver operating characteristic (ROC) curve for the triple-stage system. }
\end{figure}

The receiver operating characteristic (ROC) curve for the triple-stage system (Fig.4) characterizes the verify performance depending on how many nodes are chose for
the acceptance region. For an exemplary threshold marked with red dot the system accepts up to $97\%$ of real transient detections and rejects $~97.5\%$ of bogus candidates.
\bibliographystyle{ptapap}
\bibliography{ptabib}

\end{document}